\documentclass[doublecol,figures]{epl2}
\usepackage{amsmath,amssymb}

\title{Recommendation model based on opinion diffusion}
\author{Y.-C. Zhang\inst{1,2} \and
M. Medo\inst{1}\thanks{Email: \email{matus.medo@unifr.ch}} \and
J. Ren\inst{1} \and T. Zhou\inst{1,3} \and
T. Li\inst{2} \and F. Yang\inst{2}}
\shortauthor{Y.-C. Zhang \etal}

\institute{
  \inst{1} Physics Department, University of Fribourg,
           1700~Fribourg, Switzerland\\
  \inst{2} Department of Physics, Renmin University of China,
           Beijing 100872, PR China\\
  \inst{3} Department of Modern Physics, University of Science
           and Technology of China, Hefei 230026, PR China}


\abstract{Information overload in the modern society calls for
highly efficient recommendation algorithms. In this letter we
present a novel diffusion based recommendation model, with
users' ratings built into a transition matrix. To speed up
computation we introduce Green function method. The numerical
tests on a benchmark database show that our prediction is
superior to the standard recommendation methods.}

\begin{document}
\maketitle

\section{Introduction}
The exponential growth of the Internet \cite{Faloutsos1999} and
the World-Wide-Web \cite{Broder2000} confronts us with the
information overload: we face too many data and data sources,
making us unable to find the relevant results. As a consequence
we need automated ways to deal with the data. Recently, a lot of
work has been done in this field. The two main directions of the
research are correlation-based methods~\cite{Balab97,Pazzani99}
and spectral methods~\cite{Maslov00}. A~good overview of the
achieved results can be found
in~\cite{Herlocker04,Adomavicius05}.

Despite the amount of work done, the problem is not
satisfactorily exploited yet as both the prediction accuracy and
the computational complexity can be improved further. In this
letter we propose a~new method based on diffusion of the users'
opinions in an object-to-object network. This method can be used
for any data where users evaluate objects on an integer scale.
Using data from a real recommender application (GroupLens
project) we show that the present model performs better then
the standard recommendation methods. In addition, a Green
function method is proposed here to further reduce computation
in some cases.

\section{The model}
In the input data, the total number of users we label as $M$ and
the total number of objects as $N$ (since we focus here on the
movie recommendation, instead of the general term \emph{object}
we often use the term \emph{movie}). To make a better
distinction between these two groups, for user-related indices
we use lower case letters $i,j,k,\dots$ and for movie-related
indices we use Greek letters $\alpha,\beta,\gamma,\dots$. We
assume that users' assessments are given in the integer scale
from 1~(very bad) to 5~(very good). The rating of user $i$ for
movie $\alpha$ we denote $v_{i\alpha}$. The number of movies
rated by user $i$ we label $k_i$. The rating data can be
described by the weighted bipartite graph where the link between
user $i$ and movie $\alpha$ is formed when user $i$ has already
rated movie $\alpha$ and the link weight is $v_{i\alpha}$. Such
a bipartite graph can give rise to two different types of graphs
(often called \emph{projections}): object-to-object and
user-to-user. A~general discussion on information networks can
be found in~\cite{Newman03}, projections of bipartite graphs are
closely investigated in~\cite{Tao07,Yu07}.

\begin{figure}
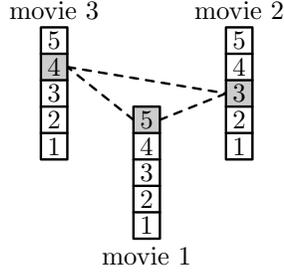

\onefigure{QDfig-1}
\caption{Graphical representation of the links created by a user
who has rated only movies 1 (rating~5), 2 (rating~3), and 3
(rating~4).}
\label{fig-links}
\end{figure}

The recommendation process starts with preparation of a
particular object-to-object projection of the input data.
Projections usually lead to a loss of information. In order to
eliminate this phenomenon, instead of merely creating a link
between two movies, we link the ratings given to this pair of
movies. As a result we obtain 25 separate connections (channels)
for each movie pair. This is illustrated in fig.~\ref{fig-links}
on an example of a user who has rated three movies; as a result,
three links are created between the given movies. When we process
data from all users, contributions from all users shall
accumulate to obtain an aggregate representation of the input
data: a weighted movie-to-movie network. From the
methodological point of view, this model is similar to the
well-known Quantum Diffusion process (see~\cite{Ian98,Kim03}).

To each user we need to assign a weight. In general, if user $i$
has rated $k_i$ movies, $k_i(k_i-1)/2$ links in the network are
created (or fortified). If we set the user weight to
$1/(k_i-1)$, the total contribution of user $i$ is directly
proportional to $k_i$, and this is a plausible
premise.\footnote{Here one can recall the famous set of
equations for PageRank $G(i)$ of webpage $i$. It has the form
$G(i)=\alpha+(1-\alpha)\sum_{j\sim i} G(j)/k_j$, where the
subscript $j$ runs over all the webpages that contain a link to
webpage $i$ ($j\sim i$), for details see~\cite{Brin98}. Here a
similar scaling of the contributions by the inverse of the node
degree arises. By a numerical solution of the set, one obtains
values $G(i)$ which are essential for the Google search
algorithm.}
Since the users who have seen only one movie add no links to the
movie-to-movie network, the divergence of the weight $1/(k_i-1)$
at $k_i=1$ is not an obstacle.

Since between each pair of movies $(\alpha,\beta)$ we create
multiple links, it is convenient to write their weights as a
$5\times5$ matrix $\mathsf{W}_{\alpha\beta}$. Each rating can be
represented by a column vector in 5-dimensional space: rating
$v_{i\alpha}=1$ we represent as
$\vect{v}_{i\alpha}=(1,0,0,0,0)^T$, rating $v_{i\alpha}=2$ as
$\vect{v}_{i\alpha}=(0,1,0,0,0)^T$, and so forth. If the vote
has not been given yet, we set
$\vect{v}_{i\alpha}=(0,0,0,0,0)^T$. Then using the linking
scheme from fig.~\ref{fig-links} and the user weights
$1/(k_i-1)$ we write
\begin{equation}
\label{W-matrix}
\mathsf{W}_{\alpha\beta}=\sum_{i=1}^M
\frac{\vect{v}_{i\alpha}\vect{v}_{i\beta}^T}{k_i-1},
\end{equation}
where we sum contributions from all users. In this way we
convert the original data represented by a weighted bipartite
graph into a weighted object-to-object network.

The non-normalized weights $\mathsf{W}_{\alpha\beta}$ form a
symmetric matrix $\mathsf{W}$ with dimensions $5N\times5N$. By
the column normalization of $\mathsf{W}$ we obtain an
unsymmetric matrix $\Omega$. It describes a diffusion process
on the underlying network with the outgoing weights from any
node in the graph normalized to unity (see also a similar
diffusion-like process in~\cite{Ou2007} and the PageRank
algorithm\footnote{Incidentally, PageRank algorithm normalizes
the flux outgoing from a node in a similar way and thus it also
represents diffusion or a random walk. If one chooses the row
normalization instead, the resulting process is equivalent to
heat conduction in the network.}).

Now we shall investigate the equation
\begin{equation}
\label{stationary}
\Omega\vect{h}=\vect{h},
\end{equation}
where $\vect{h}$ is a $5N$-dimensional vector (the first 5
elements correspond to movie 1, next 5 elements to movie 2,
etc.). Denote $n_{\alpha s}$ ($\alpha=1,\dots,M$, $s=1,\dots,5$)
the number of times movie $\alpha$ has been rated with the
rating $s$. Here we exclude the votes given by the users who
have rated only one movie because these users do not contribute
to $\Omega$. It is easy to prove that the vector
\begin{equation}
\vect{h}^*=(n_{11},\dots,n_{15},\dots,
n_{N1},\dots,n_{N5})^T
\end{equation}
is a solution of eq.~(\ref{stationary}). Moreover, the solution
is unique up to multiplication by a constant and as we will see
later, all vectors in the form $\lambda\vect{h}$,
$\lambda\neq0$, lead to identical predictions. Denote
$L:=\mathsf{1}-\Omega$ the Laplace matrix, the forementioned
uniqueness of $\vect{h}^*$ is equivalent to
$\text{rank}(L)=5N-1$, which we prove in the following
paragraph. It is worthwhile to emphasize that the unique
solution $\vect{h}^*$ reproduces some features of the original
input data, which strongly supports rationality and relevance of
the construction of $\Omega$.

Using elementary row/column operations one can shift all the
rows/columns corresponding to the zero-rows/zero-columns of
$\Omega$ to the bottom and right of $L$, leading to
$\bigl(\begin{smallmatrix}
L' & O\\
O & \mathsf{1}
\end{smallmatrix}\bigr)$, where $O$ and $\mathsf{1}$ are the
zero and the identity matrix. The dimension of $\mathsf{1}$ we
label as $D$, the dimension of $L'$ is then $5N-D$. The matrix
$L'$ has four properties: (i) All its diagonal elements are 1.
(ii) All its non-diagonal elements lie in the range $[-1,0]$.
(iii) The sum of each column is zero. (iv) In each row, there is
at least one non-diagonal nonzero element. One can prove that
the rank of any matrix with these four properties is equal to
its dimension minus one, $5N-D-1$ in this case. Since
$\text{rank}(\mathsf{1})=D$, together we have
$\text{rank}(L)=\text{rank}(L')+\text{rank}(\mathsf{1})=5N-1$.
Details of the proof will be shown in an extended paper.

The matrix $\Omega$ codes the connectivities between different
ratings in the movie-to-movie network, and could yield to a
recommendation for a particular user. Since the matrix
represents only the aggregated information, in order to
recommend for a particular user, we need to utilize opinions
expressed by this user. We do so by imposing these ratings as
fixed elements of $\vect{h}$ in eq.~(\ref{stationary}). These
fixed elements can be considered as a boundary condition of the
given diffusion process; they influence our expectations on
unexpressed ratings. In other words, large weights in $\Omega$
represent strong patterns in user ratings (\emph{e.g.} most of
those who rated movie X with 5 gave 3 to movie Y) and diffusion
of the ratings expressed by a particular user in the
movie-to-movie network makes use of these patterns.

The discussion above leads us to the equation
\begin{equation}
\label{recommendation}
\Omega_i\vect{h}_i=\vect{h}_i,
\end{equation}
where $\Omega_i:=\Omega$ for the rows corresponding to the
movies unrated by user $i$ and $\Omega_i:=\mathsf{1}$ for the
remaining rows. Such a definition keeps entries corresponding to
the movies rated by user $i$ preserved. The solution of
eq.~(\ref{recommendation}) can be numerically obtained in a
simple iterative way. We start with $\vect{h}_i^{(0)}$ where
elements corresponding to the movies rated by user $i$ are set
according to these ratings and the remaining elements are set to
zero. Then by the iteration equation
$\vect{h}_i^{(n+1)}=\Omega_i\vect{h}_i^{(n)}$ we
propagate already expressed opinions of user $i$ over the
network, eventually leading to the stationary solution
$\vect{h}_i$. Intermediate results $\vect{h}_i^{(n)}$ contain
information about the movies unrated by user $i$, which can give
rise to a recommendation. We obtain the rating prediction as the
standard weighted average. For example, if for a given movie in
$\vect{h}_i$ we obtain the 5-tuple $(0.1,0.2,0.4,0.3,0.0)^T$,
the rating  prediction is $\hat v=2.9$. Notice that if a user
has rated no movies, we have to use a different method (for
example the movie average introduced later) to make a
prediction. This feature is common for recommender systems
producing personalized predictions.

\section{Avoiding the iterations}
While simple, the iterative way to solve
eq.~(\ref{recommendation}) has one important drawback: the
iterations have to be made for every user separately.
Consequently, the computational complexity of the algorithm is
high. To get rid of this difficulty we rewrite
eq.~(\ref{recommendation}) as $L\vect{h}_i=\vect{j}_i$, again
$L=\mathsf{1}-\Omega$. Here the external flux $\vect{j}_i$ is
nonzero only for the elements representing the boundary
condition of user $i$.

The solution $\vect{h}_i$ can be formally written in the form
$\vect{h}_i=\mathsf{G}\vect{j}_i$. This resembles the well-known
Green function approach: once $\mathsf{G}$ is known,
$\vect{h}_i$ can be found by a simple matrix multiplication.
While the source term $\vect{j}_i$ is not a~priori known, we can
get rid of it by
reshuffling of the movies and grouping the boundary elements in
$\vect{h}_i$. After this formal manipulation we obtain
\begin{equation}
\label{green-start}
\binom{\vect{h}_i^\ab{B}}{\vect{h}_i^\ab{F}}=
\begin{pmatrix}
\mathsf{G}_\ab{BB} & \mathsf{G}_\ab{BF}\\
\mathsf{G}_\ab{FB} & \mathsf{G}_\ab{FF}
\end{pmatrix}
\binom{\vect{j}_i^\ab{B}}{\vect{0}},
\end{equation}
where B stands for \emph{boundary} and F for \emph{free}. Now it
follows that
$\vect{h}_i^\ab{B}=\mathsf{G}_\ab{BB}\vect{j}_i^\ab{B}$
and $\vect{h}_i^\ab{F}=\mathsf{G}_\ab{FB}\vect{j}_i^\ab{B}$,
leading us to the final result
\begin{equation}
\label{solution}
\vect{h}_i^\ab{F}=
\mathsf{G}_\ab{FB}\mathsf{G}_\ab{BB}^{-1}\vect{h}_i^\ab{B}.
\end{equation}
Since most users have rated only a small part of all $M$ movies,
the dimension of $\mathsf{G}_\ab{BB}$ is usually much smaller
than that of $\mathsf{G}$ and thus the inversion
$\mathsf{G}_\ab{BB}^{-1}$ is cheap.

The last missing point is that since $L$ is singular (as we have
mentioned, $\text{rank}(L)=5N-1$), the form of $\mathsf{G}$ can
not be obtained by inverting $L$. Hence we use the
\emph{Moore-Penrose pseudoinverse}~\cite{Penrose1955}
\begin{equation}
\label{G-form} \mathsf{G}=L^\dag=\lim_{k\to\infty}
\big[\mathsf{1}+\Omega+\Omega^2+\dots+\Omega^k-
k\vect{w}_\ab{R}\vect{w}_\ab{L}\big],
\end{equation}
where $\vect{w}_\ab{R}$ and $\vect{w}_\ab{L}$ is the right and
left eigenvector of $\Omega$ respectively, both corresponding to
the eigenvalue 1. For practical purposes, the infinite summation
in eq.~(\ref{G-form}) can be truncated at a finite value $k$.

\section{Personal polarization}
Before the described method can be used in real life examples,
there is one important technical problem. Each user has a
different style of rating---some people tend to be very strict
and on average give low marks, some people prefer to give either
1 or 5, some don't like to give low marks, and so forth. Thus,
ratings cannot be grouped together in matrices
$\mathsf{W}_{\alpha\beta}$ in the straightforward and na\"ive
way we described before for they mean different things to
different people.

To deal with this phenomenon, which we refer to as personal
polarization, \emph{unification} of ratings from different users
is used before summing users' contributions in the
object-to-object network. Consequently, before reporting
resulting predictions to a user, the output of the algorithm has
to be shifted back to the user's scale and \emph{personalization}
is needed.

To characterize the rating profile of user $i$ we use the mean
$\mu_i$ and the standard deviation $\sigma_i$ of the votes given
by him, and we compare these values with the mean $m_i$ and the
standard deviation $s_i$ of the ratings given by all users.
Notably, the quantities $m_i$ and $s_i$ take into account only
the movies rated by user $i$---if a user has a low average
rating because he has been rating only bad movies, there is no
need to manipulate his ratings. To conform a user rating profile
to the society rating profile we use the linear transformation
\begin{equation}
\label{unification}
u_{i\alpha}=m_i+(v_{i\alpha}-\mu_i)\,\frac{s_i}{\sigma_i}.
\end{equation}
Personalization of the predicted value is done by the inverse
formula $v_{i\alpha}=\mu_i+(u_{i\alpha}-m_i)\sigma_i/s_i$. We
can notice that while $v_{i\alpha}$ is an integer value,
$u_{i\alpha}$ is a real number. Nevetheless, one can obtain its
vector representation in the straightforward way: \emph{e.g.}
$u=3.7$ is modelled by the vector $(0,0,0.3,0.7,0)^T$; the
weighted mean corresponding to this vector is equal to the input
value $3.7$.

\section{Benchmark methods}
In correlation-based methods, rating correlations between users
are quantified and utilized to obtain predictions. We present
here one implementation of such a method, which serves as a
benchmark for the proposed diffusion model. The correlation
$C_{ij}$ between users $i$ and $j$ is calculated with Pearson's
formula
\begin{equation}
\label{correlation}
C_{ij}=
\frac{\sum_{\alpha}^*(v_{i\alpha}-\mu_i)(v_{j\alpha}-\mu_j)}
{\sqrt{\sum_{\alpha}^*(v_{i\alpha}-\mu_i)^2}
\sqrt{\sum_{\alpha}^*(v_{j\alpha}-\mu_j)^2}},
\end{equation}
where we sum over all movies rated by both $i$ and $j$ (to
remind this, there is a star added to the summation symbols);
$C_{ij}:=0$ when users $i$ and $j$ have no movies in common. Due
to the data sparsity, the number of user pairs with zero
correlation can be high and the resulting prediction performance
poor. To deal with this effect, in~\cite{Laureti07} it is
suggested to replace the zero correlations by the society
average of $C_{ij}$. In the numerical tests presented in this
Letter the resulting improvement was small and thus we use
eq.~(\ref{correlation}) in its original form. Finally,
the predictions are obtained using the formula
\begin{equation}
\label{corr-pred}
\hat v_{i\alpha}=\mu_i+\sum\nolimits_j'
\frac{C_{ij}}{\sum_k'C_{ik}}\,(v_{j\alpha}-\mu_j).
\end{equation}
Here we sum over the users who have rated movie $\alpha$ (prime
symbols added to sums are used to indicate this), the term
$\sum\nolimits_k' C_{ik}$ serves as a normalization factor.

As a second benchmark method we use recommendation by the movie
average (MA) where one has $\hat v_{i\alpha}=m_{\alpha}$,
$m_{\alpha}$ is the average rating of movie $\alpha$. This
method is not personalized (for a~given object, all users obtain
the same prediction) and has an inferior performance. As it is
very fast and easy to implement, it is still widely used.
Notably, when unification-personalization scheme is employed
together with MA, the predictions get personalized. As we will
see later, in this way the prediction performance is increased
considerably without a notable impact on the computation
complexity.

\section{Numerical results}
To test the proposed method based on opinion diffusion (OD) we
use the GroupLens project data, available at
\texttt{www.grouplens.org}. The total number of users is
$M=943$, the  total number of movies is $N=1\,682$, and the
ratings are integer values from 1 to 5. The number of given
ratings is 100\,000, corresponding to the voting matrix sparsity
around 6\%.

To test the described methods, randomly selected 10\% of the
available data is transfered to the probe file $\mathcal{P}$,
and the remaining 90\% is used as an input data for the
recommendation. Then we make a prediction for all entries
contained in the probe and measure the difference between the
predicted value $\hat v_{i\alpha}$ and the actual value
$v_{i\alpha}$. For an aggregate review of the prediction
performance we use two common quantities: \emph{root mean square
error} (RMSE) and \emph{mean absolute error} (MAE). They are
defined as
\begin{subequations}
\begin{eqnarray}
\label{errors}
\mathrm{MAE}&=&
\frac1n\sum_{\mathcal{P}}
\lvert v_{i\alpha}-\hat v_{i\alpha}\rvert,\\
\mathrm{RMSE}&=&
\bigg[\frac1n\sum_{\mathcal{P}}
(v_{i\alpha}-\hat v_{i\alpha})^2\bigg]^{1/2},
\end{eqnarray}
\end{subequations}
where the summations go over all user-movie pairs $(i,\alpha)$
included in the probe $\mathcal{P}$ and $n$ is the number of
these pairs in each probe dataset. To obtain a better
statistics, the described procedure can be repeated many times
with different selections of the probe data. We used 10
repetitions and in addition to the averages of MAE and RMSE we
found also standard deviations of both quantities.

\begin{figure}
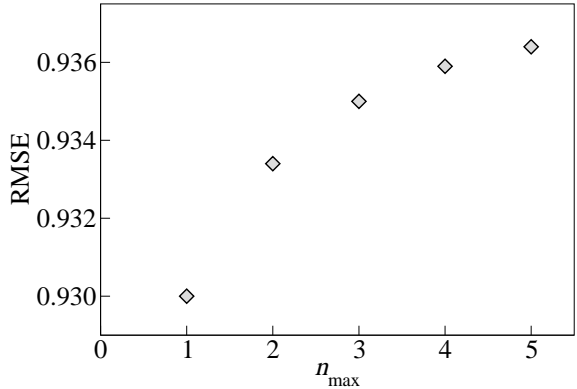

\onefigure[scale=0.3]{iterations}
\caption{Prediction performance for the predictions
$\hat v_{i\alpha}$ obtained by iterations of
eq.~(\ref{recommendation}) using various numbers of iterations
steps.}
\label{fig-iterations}
\end{figure}

In contrast with the expectations, in fig.~\ref{fig-iterations}
it can be seen that the prediction performance is getting worse
by a small amount when more than one iteration of
eq.~(\ref{recommendation}) is used to obtain the prediction.
Probably this is due to the presence of overfitting---starting
from the second iteration, our expectations are influenced not
only by actually expressed ratings but also by our expectations
about unexpressed ratings obtained in previous iteration steps.
Nevertheless, as it will be shown later, the performance
achieved by the first iteration is good and justifies validity
of the proposed model. In the following paragraphs we use only
one iteration to obtain the predictions. Consequently, the Green
function method introduced above is not necessary---we decided
to expose it in this paper because it can be useful with other
datasets.

\begin{table}
\caption{Comparison of the three recommendation methods: movie
average (MA), correlation-based method (CB), and opinion
diffusion (OD). Presented values are averages obtained using
10~different probes; standard deviations are approximately
$0.01$ in all investigated cases.}
\label{tab-comparison}
\begin{center}
\begin{tabular}{ccccc}
\hline\hline
& \multicolumn{2}{c}{no unification} &
\multicolumn{2}{c}{with unification}\\
method & RMSE & MAE & RMSE & MAE\\
\hline
MA & $1.18$ & $0.91$ & $1.01$ & $0.79$\\
CB & $1.09$ & $0.86$ & $1.09$ & $0.86$\\
OD & $1.00$ & $0.80$ & $0.93$ & $0.73$\\
\hline\hline
\end{tabular}
\end{center}
\end{table}

In table~\ref{tab-comparison} we compare the prediction accuracy
for the movie-average method (MA), the correlation-based method
(CB), and for the opinion diffusion (OD). To measure the
prediction performances we use both RMSE and MAE as defined
above. All three methods are tested both with and without
employing the unification-personalization scheme. In accordance
with expectations, for MA and OD the performances with
unification included are better than without it; for the
simplest tested method, MA, the difference is particularly
remarkable. By contrast, CB is little sensitive to the
unification procedure and when we drop the multiplication by
$\sigma_i/s_i$ from the unification-personalization process
given by eq.~(\ref{unification}), the difference disappears
completely (which can be also confirmed analytically). According
to the prediction performances shown in
table~\ref{tab-comparison} we can conclude that the diffusion
method outperforms the other two clearly in all tested cases
(RMSE/MAE, with/without unification). When computation
complexity is taken into account, it can be shown that if $M>N$,
the proposed method is more effective than correlation-based
methods (but, of course, less effective than using the movie
average).

\section{Conclusion}
We have proposed a novel recommendation method based on
diffusion of opinions expressed by a user over the
object-to-object network. Since the rating polarization effect
is present, we have suggested the unification-personalization
approach as an additional layer of the recommender system.
To allow a computation reduction with some datasets, Green
function method has been introduced. The proposed method has
been compared with two standard recommendation algorithms and it
has achieved consistently better results. Notably, it is
executable even for the large dataset (17\,770 movies, 480\,189
users) released by Netflix (a DVD rental company, see
\texttt{www.netflixprize.com}). In addition, our model is
tune-free in essence---it does not require extensive testing and
optimization to produce a high-quality output. This is a good
news for practitioners.

\acknowledgements
This work is partially supported by Swiss National Science
Foundation (project 205120-113842). We acknowledge SBF
(Switzerland) for financial support through project C05.0148
(Physics of Risk), T. Zhou acknowledges NNSFC (No.~10635040).
We kindly acknowledge the computing resources provided by Swiss
National Supercomputing Center.

\end{document}